\documentstyle[aps,epsf,prb]{revtex}
\begin{document}
\twocolumn[{
\preprint{IUCM96-26}
\title{The Quantum Hall Effect of Interacting Electrons
in a Periodic Potential}
\draft

\author{Daniela Pfannkuche$^{a,b}$ and A.\ H.\ MacDonald$^{a}$}

\address{
$^a$Department of Physics, Indiana University, Bloomington, Indiana 47405,
U.S.A.\\
$^b$Max-Planck-Institut f\"ur Festk\"orperforschung, Heisenbergstr. 1,
D-50689 Stuttgart,
Germany}

\date{Received \today}
\maketitle
\vspace*{-6ex}
\begin{abstract}
\widetext
\leftskip 54.8 pt
\rightskip 54.8 pt
We consider the influence of an external periodic potential
on the fractional quantum Hall effect of two-dimensional interacting
electron systems.  For many electrons on a torus, we find that the
splitting of incompressible ground state degeneracies by
a weak external potential diminishes as $\exp ( - L/ \xi)$
at large system size $L$.  We present numerical results
consistent with a scenario in which $\xi$ diverges
at continuous phase transitions
from fractional to integer quantum Hall states
which occur with increasing external potential strength.
%
\vspace*{2ex}
\end{abstract}
}]

\narrowtext
Theoretical studies of the quantum Hall effect (QHE) have
identified three different\cite{leshouches} sources
which can act separately or in concert to create
the charged excitation energy gap responsible for dissipationless
transport in a two-dimensional electron system (2DES).
Kinetic energy quantization is
usually primarily responsible
for the integer QHE (IQHE), while interaction
energy quantization is responsible for the
fractional QHE (FQHE).
Integer quantum Hall gaps at fractional Landau level filling factors ($\nu$)
can also arise from periodic external potentials \cite{tkkn}.
Although this instance has not yet been
realized experimentally\cite{kotthaus}, it
has played an important role in theoretical
developments, especially in giving rise to the topological
picture\cite{tkkn,topological} of the QHE.
In addition, considerations of the periodic external potential
case have cautionary\cite{cautionary} implications
for the comprehensiveness of some theoretical pictures
of the bulk QHE.

In this paper we address
the competition between electron-electron interactions
and a periodic external potential in determining the value of
the quantized Hall conductance of a 2DES in the strong magnetic
field ($B$) limit where all electrons lie in the lowest Landau level
and Landau level mixing can be neglected.  Interactions
{\it alone} give rise to charge gaps at a set of rational
values of $\nu$, all of which have odd denominators, and lead to
fractionally quantized Hall conductivity values $\sigma_{H} = (e^{2}/h) \nu$.
A periodic external potential {\it alone} splits the Landau
level into subbands in an intricate way\cite{lbands} giving rise to the
captivating `Hofstadter butterfly' illustrations
of the magnetic field dependence of the one-body spectral support.
In the non-interacting electron limit, the charge gaps
are the gaps in the one-body spectrum.
For a given unit cell area, $A_{0}$, gaps occur\cite{tkkn,amd} at
$\nu = \sigma + s A_{\phi}/ A_{0}$ where $\sigma$ and $s$ are
integers which depend in detail on the periodic
potential, and $A_{\phi}  = \Phi_{0} / B = 2 \pi \ell^{2} $
is the area penetrated by the magnetic flux quantum
$\Phi_{0} = hc/e$.  In this case the Hall conductivity
has integer quantization, $\sigma_{H} = (e^{2}/h) \sigma$.
Progress in nanolithography is beginning to allow
the fabrication of systems in which physical consequences of this
complicated electronic structure can be studied
experimentally\cite{kotthaus}, although the quantum Hall (QH) regime
has not quite been reached.   Addressing the role played by
inescapable electron-electron interactions in the QH
regime is a principle motivation for the present work.\cite{kol}
The charge gaps of the non-interacting periodic potential limit clearly
survive to
finite interaction strengths since, when all occupied subbands are
filled, the non-interacting many-particle ground
state is non-degenerate and separated from many-particle excited states by
a gap.  The weak periodic potential limit is addressed below.

The QHE of interacting electrons
in a periodic external potential is most succinctly discussed
by combining the toroidal geometry, in which quasiperiodic
boundary conditions are applied to a finite area,
with the topological picture of the quantized Hall
conductance\cite{topological}.  In this picture the
Hall conductance at zero temperature in $e^{2}/h$ units is
given by the integer valued Chern index which expresses
the adiabatic evolution of the phase of the ground state wavefunction
under cyclic evolution of the boundary condition phases.
The occurrence of a FQHE requires
the many-particle ground state to be degenerate, since in
that case the quantized Hall conductance depends on the
average Chern number of the degenerate states\cite{niu86}.
Indeed, the necessary degeneracies are a consequence\cite{haldanetrans}
of continuous translational invariance for many-electron systems on a torus
and occur for every state in the Hilbert space whether or not
there is a charge gap.

Since we restrict our attention here to
magnetic field strengths for which charge gaps occur in
both non-interacting modulated and interacting
isotropic ({\it i.e.}, zero periodic potential) limits, we
consider only rational filling factors $\nu \equiv N_{e}/N_{\phi}
= q/p$ where $p$ is odd.  (Here $N_{e}$ is the number of electrons
in a finite area ($A$) system and $N_{\phi} = A B / \Phi_{0}$ is the
number of states in the Landau level.)  A gap can be created
by a periodic potential alone at $\nu = q/p$ only if $A_{\phi}/A_{0}
= t/p $ for some integer $t$.  A weak periodic potential
will have no effect on the FQHE
provided that the ground-state degeneracy splitting which
it produces vanishes in the limit of an infinite system size
and the fractional charge gap remains finite.
In the following paragraphs we use this criterion to
demonstrate the stability of the FQHE against weak
lateral potentials.

We consider a weak periodic potential of the form
which has been most extensively investigated
for non-interacting electron systems\cite{lbands}:
$U(\vec r) = U_{0} (\cos (2 \pi x/ a_{x}) + \cos(2 \pi y / a_{y}))$.
To be compatible with quasi-periodic
boundary conditions we choose a rectangular
finite area system with sides $L_{x} = N_{x} a_{x}$ and $L_{y} = N_{y} a_{y}$,
where both $N_{x}$ and $N_{y}$ are integers,
$N_{x} N_{y} = t N_{e}/q$, and use a Landau gauge
in which the vector potential is independent of the $\hat y$
coordinate.  Our analysis is based on
three elementary properties which follow from
translational symmetry considerations \cite{haldanetrans}
in the isotropic system limit:
i) Isotropic system states can be labeled by a two-dimensional
wavevector $\vec K$ in a rectangular magnetic Brillouin-zone
with $K_{x} \in [0, L_{y}/p \ell^{2})$ and $K_{y} \in [0,L_{x} / \ell^{2})$;
ii) Pseudomomentum is conserved so that $\langle \vec K' i'|\, \rho_{\vec p}
\, | \vec K i \rangle $ is non zero only if
$\vec K' = \vec K + \vec p$;
iii) Translation by $(L_{x} q / p \ell^{2}) \hat y$ in momentum space
correspond to rigid spatial translations by
$ (2 \pi \ell^{2} / L_{y} ) \hat x$ so that
$E(\vec K,i) = E(\vec K + L_{x} \hat y / p \ell^{2},i)$
and $\langle \vec K' + L_{x} \hat y / p \ell^{2},i'| \rho_{\vec p}
| \vec K + L_{x} \hat y / p \ell^{2}, i \rangle =
\langle \vec K' i'| \rho_{\vec p} | \vec K i \rangle $ up
to a phase factor independent of $i$ and $i'$.
Here all wavevectors are understood to be reduced to the 
magnetic Brillouin-zone where necessary and $\rho_{\vec p}
\equiv \sum_{j} \exp(i \vec p \cdot \vec r_{j}) $ is the
density operator.   (The external potential contribution
to the Hamiltonian is of the form $H' = (U_{0}/2) \sum_{G} \rho_{\vec G}$.
where $\vec G$ is a nearest neighbor reciprocal lattice vector.)
When the FQHE occurs
the $p$ degenerate ground states (at $\vec K =
\vec K_{0} + m \hat{y} L_{x}/ p \ell^{2} $ with $m = 0, \cdots, p-1$)
are separated from all other eigenstates by an energy gap $\Delta$ which
remains finite in the thermodynamic limit\cite{Thouless89:12034}.

When $H'$ is included, shifts in all eigenvalues occur starting
at second order in perturbation theory.  However, it follows from
property iii) above that the shifts which appear at low
order are identical for each member of the degenerate
ground-state manifold.  Splitting of these states can result only
from terms in perturbation theory in which one member of the manifold
is coupled to another via a series of intermediate states
at energies above the gap.  It follows from property ii)
above that these terms first occur at order $M_{y}
 = [L_{x}/ p \ell^{2})/(2 \pi / a_{y}]  = [N_{x} /t] $ for
$y$-dependent terms in the external potential and at order
$M_{x} = [L_{y}/ p \ell^{2})/(2 \pi / a_{x}] = [N_{y} /t] $ for
$x$-dependent terms in the external potential.  (Here [i/j] denotes
the numerator of $i/j$ after elimination of common divisors.)
For $U_{0} \ll \Delta $ and fixed aspect ratio
the splitting of the $p$-fold degenerate
ground state manifold will therefore be
$\sim \Delta (U_{0}/\Delta)^{\min{(M_{x},M_{y})}} \sim \Delta \exp ( - L/\xi)$
where $L$ is the system size and $\xi \sim a / \ln (\Delta/ U_{0}) $.
Hence, the FQHE will survive in an infinite system
as long as $U_{0}$ is small compared to the fractional gap $\Delta$.
The exponential dependence of splitting on system size we find
is much weaker than the
Gaussian dependence ($\sim \exp ( - \nu^{2} L^{2} / 4 \ell^{2} )$)
which occurs at the lowest order of perturbation theory for
a random external potential\cite{Thouless89:12034,WenNiu90}
and is in agreement with expectations from
more heuristic arguments\cite{WenNiu90} which apply equally well to random
external potentials and are based on effective theories of the
FQHE.

\begin{figure}[bt]
   \vspace*{-2ex}
   \epsfxsize = 8.4cm
   \epsffile {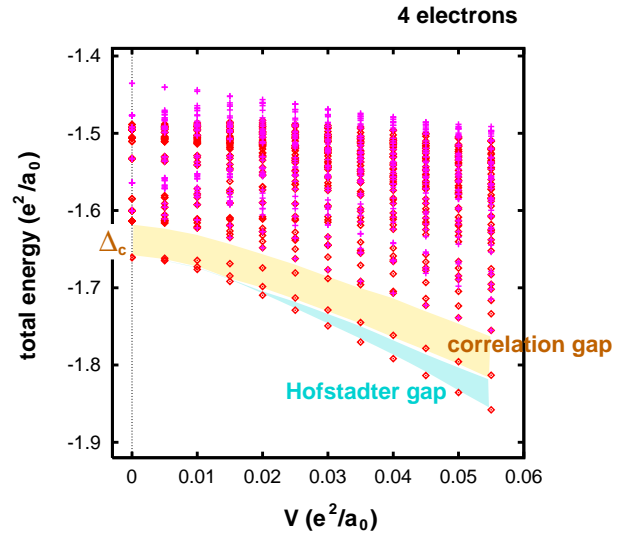}
   \caption{Low energy part of the 4-particle spectrum ($N_{x} = 2 =
N_{y}$) as a
             function of the reduced modulation strength V. Note the opening
             of the 'Hofstadter gap' $W_{1}$ at very small modulation strength
             and its coexistence with the correlation gap $W_{3,1} = E_{3}-
E_{1}$
             at a finite range of the modulation strength which are a
             consequence of the finite system size. (Different symbols
             mark different pseudomomenta.)}
   \label {fig:spectrum}
\end{figure}

This analysis suggests a scenario in which a continuous phase
transition between weak and strong periodic potential
QH states occurs at a critical value of $U_{0}$ = $U_{c}$
and that $\xi$ diverges as $U$ approaches $U_{c}$ from below.
We have attempted to test this simplest picture for
the transition between fractional and IQHEs, and to
obtain a quantitative estimate of the modulation strength at
which the putative transition occurs, by performing
numerical exact diagonalization calculations at a series of
$U_{0}$ values including Coulombic electron-electron
interactions.   Here we report results for
$\nu = 1/3$ and $A_{\phi}/A_{0}=1/3$; this case appears
to offer the greatest promise for experimental study since large
charge gaps occur in both weak and strong
modulation limits\cite{Wen93:1501}.
For these commensurability ratios, the zero temperature
Hall conductivity will change from the fractionally
quantized value  $\sigma_{H} = 1/3 \; e^{2}/h$
to the integer quantized value $\sigma_{H} = 0$ at the critical modulation
strength.  Typical finite size
results for the evolution of the low energy portion of
the spectrum with modulation potential strength are shown
in Fig.~\ref{fig:spectrum}.  We expect that
in the thermodynamic limit the splitting of
the three lowest energy states should
vanish and the correlation gap to the fourth lowest energy
state should remain finite for $U_{0} < U_{c}$.
For $U_{0} > U_{c}$ we expect the gap between the first and second lowest
energy states, which is proportional to $U_{0}$ for large $U_{0}$,
to be finite.
%
%

All the results reported here were obtained with $a_{x}=a_{y}=a$
and up to six electrons in a system with area
$A = N_{x}  N_{y} a^{2}$.
For systems with up to five electrons the
finite Hamiltonian matrix could be directly diagonalized, while for the largest
systems the low lying eigenenergies were obtained using a block Lanczos
procedure.

In interpreting these numerical results we concentrate on
the energy difference $W_{1}$
($W_{i} := E_{i}-E_{0}$, where $E_{0}$ is the ground state energy)
between the ground state and the first excited state.
Explicit perturbative calculations for the finite size 
system show that at small modulation strengths
$W_{1}$ is given at leading order in perturbation theory by
\begin{equation}
W_{1} \approx n N (V/\Delta(n,N))^{N} \Delta(n,N)/2
\label{perturbation}
\end{equation}
where $n = \min(N_{x},N_{y})$, $N = \max(N_{x},N_{y})$,
$V = U_{0} \exp(-\pi t/2 p)$ is the
external potential strength corrected by the form factor for
lowest Landau level wavefunctions, 
and $\Delta(n,N) = W_{3}(V=0)$ is the finite size correlation gap.
In the limit $V \gg \Delta$ we find that
$W_{1} = \Delta_{H} + \delta(n,N)$ where $\delta(n,N)$ is positive, and
$\Delta_{H} :=  V (3-\sqrt{3})/2$ is the gap
between the first and second subband of the Hofstadter spectrum.

\begin{figure}[tb]
   \epsfxsize = 8.3cm
   \epsffile {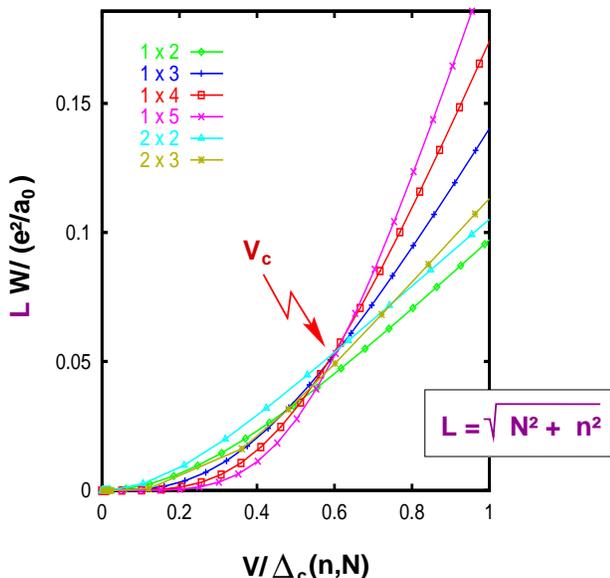}
   \caption{Gap between the ground state and the first excited state
     \protect{$W \equiv W_{1}$} scaled by the dimensionless system size
     {$L = \protect\sqrt{N_{x}^{2}+N_{y}^{2}}$} as a function of the reduced
     modulation amplitude $v = V/\Delta(n,N)$.
     Different curves represent different system sizes.
     The approximate crossing
     of these curves allows us to estimate the critical
     reduced modulation strength $v_{c}$.}
   \label {fixpoint}
\end{figure}

We see 
that the finite size numerical calculations reproduce the expected behavior
in both weak and strong modulation potential limits.
We have performed a finite size
scaling analysis\cite{sondhicqpt} to show that these results
are consistent with a continuous quantum phase transition occurring
between fractional ($\sigma_{H} = 1/3 e^{2}/h$) and integer
($\sigma_{H} = 0$) QH states at intermediate modulation strengths.
We identify the inverse gap, $W_{1}^{-1}$, with the correlation time $\xi_{T}$
beyond which the properties of the
system are sensitive to the periodic potential.
The correlation time is finite for $V_{0}$ larger than a critical
value $V_{c}$ and diverges in the thermodynamic limit
as $V_{0}$ approaches $V_{c}$ from above.
$W_{1}$ should obey the finite size scaling ansatz
\begin{equation}
 W_{1} = (L^{z})^{-1} Q(L^{1\over\nu} {{V-V_{c}}\over{V_{c}}})
\label{sc_ansatz}
\end{equation}
where $Q$ is the scaling function, $\nu$ is the correlation
length critical exponent, and $z$ is the dynamical critical exponent.
The small sizes of the systems for which
we are able to numerically solve the many-electron problem limit
the thoroughness with which we can test this ansatz.  In an
analysis which neglects aspect ratio dependence of
$W_{1}$ we use $L=\sqrt{N_{x}^{2} + N_{y}^{2}}$ as a measure of the system
size. 
Since the fractional
gap $\Delta(n,N)$ exhibits strong size and aspect ratio dependence,
we perform the scaling analysis with the modulation strength measured
in units of the finite-size gap defining $v = V_{0}/\Delta(n,N)$.

\begin{figure}[tb]
  \vspace*{-2ex}
  \epsfxsize = 8.4cm
  \epsffile {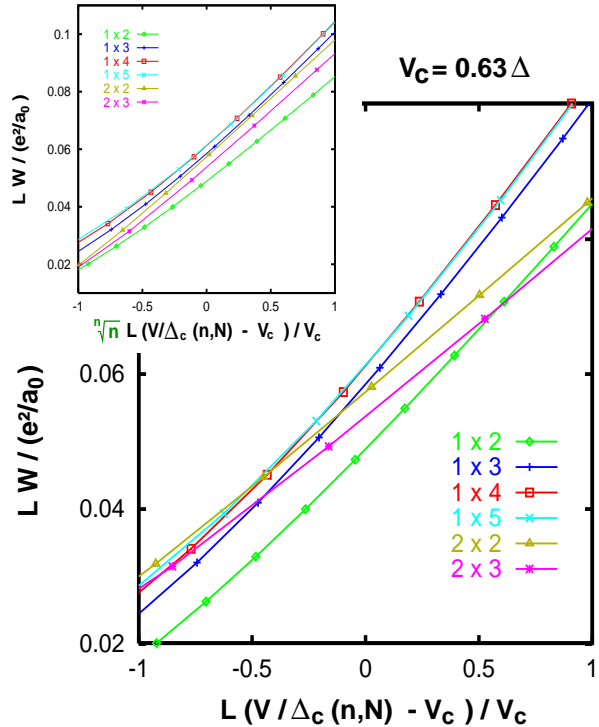}
  \caption{Scaling behavior of the energy gap near the critical point. Curves
    for the $1\times N$-system sizes are parallel, but shifted with respect to
    each other due to next-to-leading order finite size corrections. The
    $2\times N$-systems show a different slope. Inset: Non-critical
    finite-size corrections in the argument of the scaling function improve
    the scaling behavior.
    }
  \label{scaling}
\end{figure}

In Fig.~\ref{fixpoint} we plot $L W_{1}$  as a function of
$v$ for various system geometries. Assuming that $z = 1$ as in other quantum
Hall phase transitions\cite{sondhicqpt}, the approximate crossing of
these curves 
allows us to estimate the critical dimensionless modulation strength
to be $v_{c} \simeq 0.63$.
The smallness and limited range of our system sizes do not allow
us to convincingly identify the correlation length critical
exponent $\nu$.
In Fig.~3 we show the scaling function $L W_{1} = Q$
plotted as a function of $L (v-v_{c})/v_{c}$, {\it i.e.} assuming $\nu =1$.
Deviations from scaling clearly seen in these curves
become noticeably worse if $\nu$ is altered by more than $\sim 0.2$.
Data points for system sizes $N_{x} \times N_{y}  = 1 \times 4 $
and $1 \times 5$ fall on the same curve, consistent with
the scaling ansatz.  However, the curves for the $2 \times 2$ and $2 \times 3$,
systems although parallel to each other,
have a different slope than found for the $1\times N$
systems.  We can partially compensate for this large
finite size effects by assuming a non-critical size dependence
in the argument of the scaling function which is suggested by the small $V$
behavior of $W_{1}$ (see Eq.~(\ref{perturbation})).
For example, plotting $Q$ as a function of $n^{(1/n)} L (v-v_{c})/v_{c}$,
as shown in the inset of Fig.~2, improves the scaling behavior
somewhat.

Our perturbative expression for the correlation length $\xi$
suggests that the divergence of the fractional state
correlation length $\xi$ and the vanishing of
the fractional Hall gap ($\Delta = W_{3}$)
should occur simultaneously.
In our finite-size simulations we find that $\Delta$ has a minimum for
$V_{0} \sim V_{c}$, but at the system sizes we are able to study,
no compelling evidence that these minimum gaps vanish in the
thermodynamic limit emerges.

Finally we comment on the experimental implications of our work.
Broadly we emphasize that
interactions will always be important
at weak modulation strengths in any experimental
investigation of a `Hofstadter butterfly' system.
More particularly, we propose
that continuous phase transitions between integer and
FQH ground states will frequently
occur as modulation strengths are weakened.
Numerical exact diagonalization calculations for
very small systems, while not able to offer
compelling evidence, are consistent with this suggestion.
For the case of $\nu=1/3$ and one electron per unit cell
of the periodic potential we estimate that the phase
transition will occur when the ratio of the modulation strength to
the fractional Hall gap is $\sim 0.7$.  $T \ne 0$ finite-temperature scaling
analysis of the phase transition\cite{sondhicqpt} implies
that the Hall conductivity will change from
$\sigma_{H} = 1/3 e^{2}/h$, to $\sigma_{H} = 0$
over an interval of modulation strength which
vanishes as $T^{1 \over \nu z}$, becoming increasingly sharp as
the temperature is lowered.  If our limited finite size data is indicative,
$\nu z \sim 1$ rather than $\sim 0.4$ as found experimentally
in field driven integer quantum Hall transitions\cite{hpwei}.
Simultaneously, the
longitudinal conductivity will change from $\sigma_{l} = 0$ (corresponding to
dissipationless transport when the fractional gap separates the ground state
from the excited states) to a finite value at the transition point and back to
$\sigma_{l} = 0$ when the Hofstadter gap opens at modulation
strengths larger than $V_{c}$.  These behaviors would signal
the continuous quantum phase transition envisaged here.
For a system with a carrier density $n = 4 \times 10^{10} {\rm cm}^{-2}$ the
commensurability ratios studied here
would be realized with a potential period
$a \sim 50 {\rm nm}$ and a field $B \sim 5 {\rm Tesla}$.
We would then predict that the transition between integer and fractional
QHEs would occur for a modulation strength
$\sim 0.1 {\rm meV}$.  These periods, modulation strengths,
and fields are not far removed from what can be achieved with
current lithographic technology.  The physics of the
Hofstadter butterfly QH system, when it is finally realized,
will be greatly enriched by electron-electron interaction effects.

This work was supported in part by the National
Science Foundation under grant DMR-9416906.  D.P. appreciates financial
support by the Deutsche Forschungsgemeinschaft. The authors are
grateful for helpful interactions with Steve Girvin, Qian Niu, Erik
S{\o}rensen and Ulrich Z\"{u}licke.

\vspace*{-4ex}



\end{document}